\documentstyle[preprint,prl,aps,epsfig]{revtex}
\begin{document}
\def\sn2{$\sin^22\theta$}
\def\dm2{$\Delta m^2$}
\def\ch2{$\chi^2$}
\def\n{$\nu~$}
\def\nue{$\nu_e~$}
\def\nus{$\nu_s~$}
\def\numu{$\nu_{\mu}~$}
\def\nutau{$\nu_{\tau}~$}
\def\ar{$\rightarrow~$}
\def\lrar{$\leftrightarrow~$}
\def\ltap{\ \raisebox{-.4ex}{\rlap{$\sim$}} \raisebox{.4ex}{$<$}\ }
\def\gtap{\ \raisebox{-.4ex}{\rlap{$\sim$}} \raisebox{.4ex}{$>$}\ }
\draft
\begin{titlepage}
\preprint{\vbox{\baselineskip 10pt{
\hbox{ }
\hbox{ }
\hbox{}}}}
\vskip -0.4cm
\title{ \bf Neutrino Mass Spectrum with $\nu_{\mu}$ $\rightarrow$
$\nu_s$ Oscillations of Atmospheric Neutrinos}
\author{Q.Y. Liu$^{1)},~$  A.Yu. Smirnov$^{2,3)}$}
\address{1) Scuola Internazionale Superiore di Studi Avanzati, I-34013
Trieste, Italy}
\address{2) Abdus Salam International Centre for Theoretical Physics,
I-34100
Trieste, Italy}
\address{3) Institute for Nuclear Research, Russian Academy of Sciences, 
107370 Moscow, Russia}
\maketitle
\begin{abstract}
\begin{minipage}{5in}
\baselineskip 16pt
We consider the ``standard'' spectrum of the active 
neutrinos (characterized by strong 
mass hierarchy and small mixing) with 
additional sterile neutrino, $\nu_s$. 
The sterile neutrino  mixes strongly  with the muon 
neutrino, so that $\nu_{\mu} \leftrightarrow \nu_s$ 
oscillations solve the atmospheric neutrino problem. 
We show that the parametric enhancement of  
the $\nu_{\mu} \leftrightarrow \nu_s$ oscillations occurs for the 
high energy atmospheric neutrinos which cross the core of the Earth. 
This  can be relevant for the anomaly observed by  
the MACRO experiment.  
Solar neutrinos are converted both to $\nu_{\mu}$ and
$\nu_s$. The heaviest neutrino ($\approx \nu_{\tau}$) may compose the hot
dark matter of the Universe. Phenomenology of this
scenario is elaborated and crucial experimental signatures are
identified. We also discuss properties of the underlying neutrino mass
matrix.  
 
\end{minipage}
\end{abstract}
\end{titlepage}
\newpage


\def\C{C}
\def\S{S}

\setcounter{footnote}{0}

\vglue 1.5cm
\leftline{\bf 1. Introduction}
\vskip 0.2cm
\indent 
Reconstruction of the neutrino mass spectrum and  
lepton mixing is one of the fundamental problems 
of  particle physics. This problem 
may be solved with  new generation of the neutrino experiments. 

It is believed that the most natural scenario of the neutrino mass 
and mixing is the one with strong mass hierarchy 
\begin{equation}
m_3 = (1 - 5)~ {\rm eV}, ~~~~ 
m_2 = (2-3) \cdot 10^{-3}~ {\rm eV},~~~~  
m_1 \ll m_2  
\label{mass}
\end{equation}
and small lepton mixing 
(comparable
with mixing in the quark sector),  so that 
$\nu_e \approx \nu_1$,  
$\nu_{\mu} \approx \nu_2$,  
$\nu_{\tau} \approx \nu_3$.  
In this scenario the heaviest neutrino,   
$\nu_3$,  composes the hot dark matter (HDM)  of the Universe, while  
the $\nu_e \rightarrow \nu_{\mu}$ resonance conversion solves the 
solar neutrino ($\nu_{\odot}$) - problem \cite{Alexei1}. This scenario 
(which can be called the ``standard scenario") 
follows naturally from the see-saw mechanism \cite{SW1,SW2} with the neutrino 
Dirac mass matrix similar to the quark mass matrix: 
$m^D_{\nu}~\sim~ m_{up}$ and  the Majorana masses  of
the right handed neutrinos at the intermediate mass scale:    
 $10^{10}-10^{13}$ GeV \cite{Alexei2}. 
Short length $\bar{\nu}_e \leftrightarrow \bar{\nu}_{\mu}$ and 
${\nu}_e \leftrightarrow {\nu}_{\mu}$ oscillations proceed via mixing of
$\nu_e$ and $\nu_{\mu}$ in the heaviest state $\nu_3$ 
\cite{indirect}. The transition probability for the LSND
experiment \cite{lsnd}  
is small, $P \lesssim 10^{-3}$; It can be reconciled 
with data at $3\sigma$ level only \cite{BandG}.

In the standard scenario, there is no oscillation 
solution of the atmospheric neutrino ($\nu _{atm}$) - problem \cite{atm}. 
Recently, the Super-Kamiokande \cite{SK} 
and  Soudan \cite{Soudan} experiments have confirmed 
the existence of the problem. Moreover, observations of the 
zenith angle as well as $L/E$ (distance/energy) 
dependencies of the muon neutrino 
deficit strongly indicate an oscillation solution of the problem.

One can  accommodate the oscillation solution of the
$\nu_{atm}$- problem  modifying the  standard scenario.
A straightforward possibility  is to diminish  
$m_3$ down to $\sim$ (0.03 - 0.3) eV 
implied by the  $\nu_{atm}$- data.  
For $m_3 \sim  (0.5 - 0.6)$ eV 
one can try also to explain the LSND result 
 \cite{Fuller} ignoring the zenith angle dependence in the 
atmospheric neutrino deficit.  
Further modification is needed to get sufficient amount of 
the hot dark matter in the Universe. 
In  this connection  the  degenerate mass spectrum with 
$m_1~\approx~m_2~\approx m_3~\sim 1 - 2$ eV 
has been suggested \cite{cald}.  
Simultaneous explanation of the LSND result and the zenith angle 
dependence in the atmospheric neutrinos requires 
depart from the three neutrino scheme.

In this paper we will consider a possibility to rescue the 
``standard scenario". The idea is to  keep 
the standard structure with strong mass hierarchy (\ref{mass}) 
and  small mixing for active neutrinos and explain the 
atmospheric neutrino problem by oscillations of  
$\nu_{\mu}$ to new neutrino state  $\nu_s$:   
$\nu_{\mu}  \leftrightarrow \nu_s$ \cite{Akhmedov}.   
This new neutrino should be a 
sterile neutrino (singlet of $SU_2 \times U_1$) in order to
satisfy the bound on the number of neutrino species 
from the $Z^0$-decay width. Large $\nu_s - \nu_{\mu}$ mixing
can be related to a singlet character of $\nu_s$. 

The $\nu_{\mu}  \leftrightarrow \nu_s$ oscillations 
as a solution of the atmospheric neutrino problem 
(see {\it e.g.} \cite{Akhmedov}) 
have been considered previously in the context of the model with mirror 
symmetry \cite{FootandVol}. In contrast with \cite{FootandVol} we
introduce only one singlet state, and   
phenomenology of our scenario substantially differs from that in 
\cite{FootandVol}. 

As we will show, in the suggested scenario
the zenith angle  dependence in the atmospheric neutrinos
can be reconciled with
LSND result at $3\sigma$ level ($P \sim 10^{-3}$).
For relatively large mass of the new neutrino the predicted LSND
probability can be enhanced up to the experimental
value $P \sim 3 \cdot 10^{-3}$, however in this case
no zenith angle dependence is expected.

The paper is organized
in the following way.  In sect. 2 we describe the scenario. Then we
consider predictions for the atmospheric 
(sect. 3) and solar  (sect. 4) neutrinos, for the LSND experiment
(sect. 5) and neutrinos from supernovas (sect. 6). Implications 
to the primordial nucleosynthesis are
discussed in sect. 7. We comment on properties of the underlying 
mass matrix in sect. 8. 

 
\vskip 0.5cm

\leftline{\bf 2. Scenario}
\vskip 0.2cm
The level and mixing scheme is shown in (fig. 1). 
The mass states $\nu_1$, $\nu_2$, $\nu_3$ have the hierarchy 
of the standard scenario (\ref{mass}).  
The mass of the  additional  state $\nu_4$  equals  
$m_4 \sim (0.5~-~7) \cdot 10^{-1}$ eV, 
so that $\Delta m^2_{24}~\sim~m^2_4$ is in
the range of solution of the $\nu_{atm}$-problem. 
The  singlet neutrino,  $\nu_s$,  and the muon neutrino, $\nu_{\mu}$,  
mix strongly in the mass states $\nu_2$ and $\nu_4$. 
We define  the elements of the  mixing matrix $U_{\alpha i}$  
which relate the flavor $\nu_{\alpha} 
\equiv (\nu_e,\nu_{\mu},\nu_{\tau},\nu_s)$ and the mass 
$\nu_i \equiv (\nu_1,\nu_2,\nu_3,\nu_4)$ states   
as:
$$ 
\nu_{\alpha} ~=~U_{\alpha i} \nu_i ~,~~~~~~~
\alpha =e,\mu,\tau,s~; ~~~~i=1,2,3,4.
$$
All mixings except for $\nu_{\mu}$ - $\nu_s$ are small: 
$U_{\mu 2} \sim U_{\mu 4}\sim U_{s 2} \sim U_{s 4} = 0.5 - 0.8$, 
and  $U_{\alpha, i} \ll 1$ for others.  


\vskip 0.3cm
\leftline{\bf 3. Atmospheric Neutrinos }
\vskip 0.2cm
A solution of the atmospheric neutrino problem is based, mainly, on
$\nu_{\mu} \leftrightarrow \nu_s$ oscillations.  
The muon neutrino state can be written as:
\begin{equation}
\nu_{\mu}~=~\sqrt{1-U_{\mu3}^2 - U_{\mu4}^2} \cdot \nu'~ +~U_{\mu3}
\cdot \nu_3~+~U_{\mu4} \cdot \nu_4 ~, 
\label{atmstate}
\end{equation}
where $\nu'$ is the combination of the light 
states $\nu_1$ and $\nu_2$. 
According to our scenario the  admixtures of the $\nu_e$  in the heavy 
states $\nu_3$ and  $\nu_4$  are small:  
$U_{e 3}, U_{e 4} \ll 1$.   
In fact, these matrix elements  are  strongly
restricted by recent CHOOZ result \cite{CHOOZ}.   
For small $U_{e 3}$ and  $U_{e 4}$    
 the oscillation effects  related to 
splitting between the states $\nu_1$ and $\nu_2$  
are negligibly small. 
Note that splitting between these states in matter can be
large enough, so that the oscillation length is smaller than 
the diameter of the earth. 
However, the  effective mixing angle is strongly suppressed by matter. 
(The mixing may not be suppressed if 
$U_{e 3}$ and  $U_{e 4}$ are large. 
In fact,   $U_{e 4}$ can be large,  if $m_3 < 3 \cdot 10^{-2}$ eV. 
This possibility will be described elsewhere \cite{LisiAlexei}).  
Thus, the task is reduced to oscillations in the 
$(\nu', \nu_3, \nu_4)$ system (see (\ref{atmstate})) and the mixing
elements 
$U_{\mu 3}$,  $U_{\mu 4}$ are relevant only.  
Being in the eV - range, the state $\nu_3$ produces
just the averaged oscillation effect, so that   
 the $\nu_{\mu}$ survival probability can be written as 
$$
 P(\nu_{\mu} \rightarrow \nu_{\mu}) = (1-|U_{\mu3}|^2)^2 
P_2 + |U_{\mu 3}|^4~. 
$$
Here $P_2$ is the survival probability for two neutrino oscillations
in matter with parameters 
$\Delta m^2_{42}$ and 
\begin{equation}
\sin^2 2\theta_{atm} =
\frac{4|U_{\mu 4}|^2 (1 - |U_{\mu 3}|^2 - |U_{\mu 4}|^2)}
{(1 - |U_{\mu 3}|^2)^2}~. 
\label{atmangle}
\end{equation}
These oscillations are described by
the effective Hamiltonian
$$
H ~=~\left ( \begin{array}{cc}
0~~ & \displaystyle { \Delta m^2_{42} \over 4E} \sin 2\theta_{atm} \\ 
\displaystyle { \Delta m^2_{42} \over 4E} \sin 2\theta_{atm} &
~~~~~~~ \displaystyle { \Delta m^2_{42} \over 2E} \cos 2\theta_{atm} + 
V
 \end{array} \right ),
$$
with the matter potential 
$$
V~=~\pm \sqrt{2} G_F \displaystyle { \left [ {1 \over 2} N_n~ +~
(1-k)(N_e - {1 \over 2} N_n ) \right ] }~,  
$$
where 
$
k \approx  
 |U_{s4}|^2 / (|U_{s 4}|^2 + |U_{e 4}|^2 ) \leq 1 $.  
(The minus sign of $V$ is for anti-neutrino.) $N_n(t)$ and $N_e(t)$ are 
the neutron and the electron number densities correspondingly. 
In the pure $\nu_{\mu} \leftrightarrow \nu_s$ case one has  $k=1$, 
and $V =  \sqrt{2} G_F N_n/2$, so that the neutron density 
contributes only. 
With decrease of $k$  the potential increases. 

If $\Delta m^2_{42} \le 10^{-2}$ eV$^2$, the oscillations are not
averaged and $P_2$ leads to the zenith angle dependence of the
$\nu_{\mu}$ - deficit. For large $U_{\mu 3}$ the total effect is the
combination of the averaged $\nu_{\mu} \leftrightarrow \nu_{\tau}$
oscillations and non-averaged $\nu_{\mu} \leftrightarrow \nu_s$
oscillations. 

The double ratio equals to survival probability:
\begin{equation} 
R~\equiv~\displaystyle { {(\mu / e)_{osc}} \over {(\mu /
e)_{MC}}}~=~P(\nu_{\mu} ~\leftrightarrow~ \nu_{\mu}) ,  
\label{douR}
\end{equation} 
as in the case of $\nu_{\mu} \leftrightarrow \nu_{\tau}$ oscillation. 
In (\ref{douR}) ${(\mu / e)_{osc}}$ and ${(\mu /
e)_{MC}}$ are the ratios of numbers 
of the $\mu$-like to $e$-like events in presence
of oscillations and without oscillations correspondingly.

Difference between the
$\nu_{\mu} \leftrightarrow \nu_{\tau}$ and 
$\nu_{\mu} \leftrightarrow \nu_s$  channels can follow from 
matter effect which influences the latter channel only. 
For maximal mixing the oscillation length $l_m$
and the effective mixing $\theta_m$ which 
determines the depth of  oscillations can be written as 
\begin{equation}
l_m = \frac{2\pi}{V}\left[ 1 + \xi  \right]^{-1/2}, ~~~ 
\sin^2 2\theta_m = \frac{\xi}{1 + \xi}, ~~~~ 
\xi \equiv \left[\frac{\Delta m^2}{2 E V}\right]^2.  
\label{param}
\end{equation}
For $\Delta m^2_{24} > 4 \cdot 10^{-3}$ eV$^2$ both for 
the sub-GeV and for multi-GeV ($E \sim 3 - 5$ GeV) 
events  we get $\xi > 10$ and the 
matter effect is negligibly small ($ < 10 \%$ in probability).  
Thus, in significant region of parameters 
(masses and mixing) at low energies 
the $\nu_{\mu} \leftrightarrow \nu_s$ oscillations reproduce  
results of the  
$\nu_{\mu} \leftrightarrow \nu_{\tau}$ oscillations, 
as far as the charged current ($CC$) interactions are concerned. 
(The number of the tau-lepton events is small.)

The matter effect  can   be important,  
for multi-GeV events,  if  $\Delta m^2_{24} < 3 \cdot 10^{-3}$ eV$^2$.  
In particular,  the zenith angle dependence of ratios 
can be modified. 
Since  matter suppresses the $\nu_{\mu} \leftrightarrow \nu_s$ 
oscillations, one expects weaker oscillation effect for the 
upward-going events, and therefore, 
flattening of the  zenith angle,  $\Theta$, dependence. 
In particular,  
$R(\cos \Theta = -1) ~>~1/2$ in the $\nu_{\mu} \leftrightarrow 
\nu_s$  case, whereas for the 
$\nu_{\mu}$ $\leftrightarrow$ $\nu_{\tau}$ case   
one gets $R(\cos \Theta = -1) ~=~1/2$ provided the 
mixing is maximal.

The matter effect becomes important for  
through-going and stopping muons \cite{Akhmedov} 
produced by high energy neutrinos  
even for large $\Delta m^2_{24}$. 
With increase of energy (decrease of $\Delta m^2$) $\xi \rightarrow 0$ 
and the oscillation length increases approaching the 
asymptotic value  determined by 
the potential only: $l_m \approx 2 \pi/V $. 
In the same time, the effective mixing angle decreases, so that 
the oscillation effects become weaker. 

The zenith angle dependence of the survival probability 
for fixed energies is shown in fig. 2. 
According to fig. 2  the dependence 
has two dips: the wide dip with minimum at  
$\cos \Theta \sim (- 0.5$  -- $ - 0.4)$ and the narrow one with minimum at 
$\cos \Theta \sim - 0.9$. 
With increase of energy the depths  of dips 
(peaks in the transition probability) decreases, 
whereas the shape of dips,  
and in particular, positions of minima change weakly.  
This behavior can be understood  in the following way. 
The positions of minima and maxima are 
determined by the phase of oscillations:  
\begin{equation}
\Phi = 2 \pi \int \frac{dL}{l_m} \approx \int d L V ~.  
\end{equation} 
For $E > 20$ GeV  and $\Delta m^2 < 5 \cdot  10^{-3}$ eV$^2$ 
we have  $\xi  < 0.2$,  so that  the length of oscillations 
and therefore the phase are  
determined  by the potential $V$ and only weakly 
depend on energy.  As a consequence,  the phase of oscillations is 
fixed  by the zenith angle. 
According to the model 
of the Earth \cite{earth}, the trajectories touch the core at 
$\cos \Theta \approx - 0.8$.  
For 
$\cos \Theta >  - 0.8$ 
neutrinos cross the mantle only, and 
the wide dip  at $\cos \Theta >  - 0.8$ 
is due  to the oscillations in the mantle. 
At  $\cos \Theta \approx - 0.4$      
the phase $\Phi$ equals $\pi$ 
which corresponds to the minimum of $P$. 
It turns out that at $\cos \Theta \approx - 0.8$ 
the phase is  $\Phi = 2 \pi$ and the oscillation effect is zero. 
 
At $\cos \Theta < - 0.8$ neutrinos cross both 
the mantle and the core. That is,     
the narrow dip   
is due to oscillations of neutrinos whose trajectories 
cross the core.  In this dip  the oscillation effect  is 
even stronger than in the wide dip, in spite of larger  
 density of the core. 
This dip (or peak in the transition probability)  is  
the parametric resonance peak \cite{param,param2}:   
the parametric enhancement of the 
$\nu_{\mu} \leftrightarrow \nu_s$  oscillations 
occurs for neutrinos which cross the core of the Earth. 
Indeed, these neutrinos cross three layers with approximately 
constant densities: the mantle, the  core, and again the mantle, 
divided by sharp change of density on the borders.  
Let us denote the phases acquired in these layers by   
$\Phi_{m1}$,  $\Phi_{c}$,   $\Phi_{m2}$  
(obviously, $\Phi_{m1} =  \Phi_{m2}$).   
It turns out that for 
$\cos \Theta \sim - 0.88$ 
the following equality is realized: 
\begin{equation}
\Phi_{m1} \approx \Phi_{c} \approx \Phi_{m2} \approx \pi~.  
\label{equality}
\end{equation} 
That is, for  each layer the size of the  layer  coincides with 
half of the averaged oscillation length: 
$\bar{l}_{m i}/2 \approx L_i $. This is 
nothing but the condition of the  
parametric resonance \cite{param,param2}. The amplitude  
of oscillations is enhanced. Maximal transition probability 
which can be achieved in the  case of three layers equals: 
\begin{equation} 
P_{tr} = \sin^2 (4 \theta_m^{mant} - 2\theta_m^{core}),  
\label{probab}
\end{equation}
where $\theta_m^{mant}$ and $\theta_m^{core}$ are the averaged 
mixing angles in the mantle and in the core correspondingly.  
 Let us stress that for sufficiently large energies 
the equality (\ref{equality}) does not 
depend on neutrino masses. The equality  is determined basically 
by the density distribution in the Earth and by the  
potential which in turn is fixed by the channel of oscillations 
and the Standard Model interactions.  
The equality is fulfilled for oscillations into 
sterile neutrinos only. For the $\nu_{\mu} \leftrightarrow \nu_{e}$
channel the  
potential is two times larger and (\ref{equality}) is failed.   
Note that for $\cos \Theta = - 1$ the 
 oscillation effect is small.

Qualitatively  the zenith angle dependence shown in fig. 2 
is similar to that  observed in the MACRO 
experiment \cite{MACRO}. There are two  dips 
with minima at  
$\cos \Theta =  (- 0.4$ --- $- 0.6)$ and 
$\cos \Theta = (- 1.0$ --- $- 0.8$). In our interpretation 
the latter dip could be 
due to the parametric enhancement of 
the $\nu_{\mu} \leftrightarrow \nu_s$ 
oscillations of neutrinos which cross both the mantle and the 
core of the Earth. 
Similar features, although not so profound, have been observed in the  
Baksan experiment \cite{BAKSAN} which has 
energies  of the detected neutrinos 
similar to those in the  MACRO experiment.  
In the Super-Kamiokande experiment the effect 
is expected to be weaker,  since the energy threshold 
of the through-going  muon detection and corresponding  energies  of 
neutrinos are higher \cite{SK}. Detailed comparison of the effect with 
data  will be given elsewhere \cite{LMS}.

Another consequence of the $\nu_{\mu} \leftrightarrow \nu_s$ 
oscillations is that the region
excluded by the IMB analysis of the   
stopping and through-going muons is shifted  by factor $\sim 3$  
to larger 
values of $\Delta m^2$ \cite{Akhmedov}. This allows one to reconcile 
the excluded IMB region with the preferable  Super-Kamiokande domain.

Crucial check of the $\nu_{\mu} \leftrightarrow \nu_s$ 
oscillations can
be obtained from studies of  events produced by the neutral current
($NC$-) interactions  of the atmospheric neutrinos.  The number of
the $NC$-events is suppressed by the 
$\nu_{\mu} \leftrightarrow \nu_s$ oscillations and 
unchanged by the $\nu_{\mu} \leftrightarrow \nu_{\tau}$ oscillations. 
It is difficult to detect the  elastic neutral current interactions.  
In this connection, it was 
proposed  to study $\pi^0$'s produced, mainly, by the neutral currents 
\cite{K2K,VS}:
$$
\nu~N~\rightarrow~ \nu~\pi^0~N~  
$$
(in \cite{K2K} the long base line experiment K2K was discussed).
Subsequent decay $\pi^0$ $\rightarrow$ $\gamma \gamma$ is 
identified as the two showering event with certain invariant mass. 
Practically, one can measure the number of 
$\pi^0$-events, $N_{\pi^0}$,   
and find the ratios $N_{\pi^0}/ N_e$ and  $N_{\pi^0}/ N_{\mu}$,  
where $N_e$ and   $N_{\mu}$ are the numbers of 
$e$-like and  $\mu$-like events induced, mainly, by the charged currents. 
This eliminates  uncertainties related to  absolute values of
the atmospheric neutrino fluxes. Another proposal \cite{VS}  
 is to measure the ratio of $N_{\pi^0}$ and  
the number of events with 1 $\pi$-production by the charged
currents. The latter dominate in the two-prong (ring) events sample, 
$N_{2ring}$, and experimentally it is possible to  determine the ratio  
\begin{equation} 
R_{\pi^0 / 2ring}~\equiv~\displaystyle { N_{\pi^0} \over N_{2ring}}~. 
\label{rpitwo}
\end{equation} 
This allows one to diminish  uncertainties related to 
cross-sections. Also one can study the ratio 
$ N_{\pi^0}/ N_{multi-ring}$.  
Since $\nu_{\mu} \leftrightarrow \nu_s$  oscillations suppress 
equally the $CC$- and $NC$- interactions of the muon neutrinos, while 
the $\nu_{\mu} \leftrightarrow \nu_{\tau}$ 
oscillations suppress the $CC$- events only,
one expects
$$
R_{\pi^0 / multi-ring}(\nu_{\mu} \rightarrow \nu_s)~ 
< ~ R_{\pi^0 / multi-ring}(\nu_{\mu} \rightarrow \nu_{\tau})~.
$$ 
Similar inequality is expected for $ N_{\pi^0}/ N_e$.
For the survival probability $P \approx 0.6$ the difference can be
as big as $(30-50)\%$ \cite{VS}. 
Thus with about 300 $\pi^0$-events which will
be detected by the Super-Kamiokande experiment within  2 - 3  years one
will be able to disentangle  solutions. Additional  check 
will be done by the long baseline experiment K2K (KEK to Super Kamiokande)
\cite{K2K}. In particular, one will be able
to compare the number of $\pi^0$'s produced in the front (close to
neutrino production target) detector  
and in the Super-Kamiokande detector. 
However, the number of $\pi^0$ expected in the SuperKamiokande is 
expected to be small 
( $\sim 30$ ).


\vskip 0.5cm
\leftline{\bf 4. Solar Neutrinos }
\vskip 0.2cm
\indent Due to strong mass hierarchy, a dynamics of conversion of 
the solar
neutrinos is reduced to one $\Delta m^2$ task. Indeed, the electron
neutrino state can be written as 
\begin{equation}
\nu_e = \cos \phi \cdot \nu_e' + \sin \phi \cdot \nu_H~, 
\label{}
\end{equation}
where 
\begin{equation}
\sin \phi  \equiv \displaystyle{\sqrt{U_{e3}^2 + U_{e4}^2}}~,   
\label{}
\end{equation}
and $\nu_e'$ is the combination of the light states:
\begin{equation}
\nu_e' = \cos \omega  \cdot \nu_1 + \sin \omega  \cdot \nu_2 
\label{}
\end{equation}	  
with 
\begin{equation}
\cos \omega = \displaystyle {U_{e1} \over {\sqrt{U_{e1}^2 +
U_{e2}^2}}}~.  
\label{}
\end{equation}
The $\nu_H$ is a combination of the heavy states:
\begin{equation}
\nu_H = \displaystyle {{U_{e3}  \cdot \nu_3 + U_{e4}  \cdot \nu_4} \over
{\sqrt{U_{e3}^2 + U_{e4}^2}}}~.  
\label{}
\end{equation}
For evolution of the solar neutrinos produced as $\nu_e$ we get the
following picture. The states $\nu_3$ and $\nu_4$ decouple from the
system, thus leading to the averaged oscillations result. The state
$\nu_e'$
converts resonantly to its orthogonal state:
\begin{equation}
\nu_x = \cos \omega  \cdot \nu_2 - \sin \omega  \cdot \nu_1 ~. 
\label{}
\end{equation} 
Using this picture we find that the survival probability for the solar 
$\nu_e$ can be written as 
\begin{equation} 
P(\nu_e \rightarrow \nu_e) = \cos^4 \phi ~
P_{2MSW} + |U_{e4}|^4+|U_{e3}|^4~ .~~ 
\label{Ps_sur}
\end{equation}
(Here we also take into account that oscillation effects in
$\nu_3 - \nu_4$ system are averaged out.) $P_{2MSW}$ is the  
$\nu_e'$ survival probability of $\nu_e'$ $\rightarrow$ 
$\nu_x$ resonance conversion. The amplitude of the probability 
can be obtained by solving  the evolution equation  
\begin{equation}
i {\displaystyle d \over \displaystyle dt}  \nu =~ H_2 \nu 
\label{}
\end{equation}
for  the two neutrino 
system $ \nu  \equiv (\nu_e',  \nu_x)^T$.   
The effective Hamiltonian
\begin{equation} 
H_2 = \left ( \begin{array}{cc}
0 & \displaystyle { {\Delta m^2_{21} \over 4E}{\sin 2\omega }} -
{\displaystyle{1 \over 4}} U_{e4} \sin 2 \psi \cdot
 {\displaystyle {\sqrt{2} G_F N_n }} \\ 
& \\ 
\displaystyle { {\Delta m^2_{21} \over 4E} {\sin 2\omega }} - 
{\displaystyle {1 \over 4}} U_{e4} \sin 2 \psi  \cdot 
 {\sqrt{2} G_F N_n } & \displaystyle { {\Delta m^2_{21} \over 2E} \cos
2\omega - \sqrt{2} G_F N^{eff}} \end{array} \right )
\label{evolution}
\end{equation}
can be found from the whole 4-neutrino Hamiltonian. 
In (~\ref{evolution}) $\Delta m^2_{21} \equiv  m_2^2-m_1^2$, 
and $\psi$ is determined by   
\begin{equation}
\sin^2 \psi = {\displaystyle |U_{\mu4}|^2 \over {\displaystyle
|U_{\mu4}|^2+|U_{s4}|^2}}~.  
\label{}
\end{equation}
Note that for small $U_{\mu 3 }$ the angle   
$\psi$ coincides practically with $\theta_{atm}$ (\ref{atmangle}) 
responsible for oscillations of the atmospheric 
neutrinos. In (\ref{evolution}) the effective density equals  
\begin{equation} 
N^{eff} = 
N_e\cos^2 \phi - {1\over {2}} N_n(\sin^2 \psi
   - |U_{e4}|^2 \cos^2 \psi ). 
\label{solNeff}
\end{equation}  
If mixing of $\nu_e$ in heavy states is negligibly small,   
the whole survival probability is reduced to
$P_{2MSW}$ according to (\ref{Ps_sur}).\\

In our scenario new features of the matter effect appear 
in comparison with pure active ($\nu_e$ $\rightarrow$ $\nu_{\mu}$) or
pure sterile ($\nu_e$ $\rightarrow$ $\nu_s$) neutrino
conversions. In particular, for non-zero admixture of the $\nu_e$ - flavor
in the  state $\nu_4$ ($U_{e4} \neq 0$) the off-diagonal elements of
the  Hamiltonian (\ref{evolution}) 
depend on matter density, so that  
the  mixing can be induced, at least partly,  by the matter
effect. 
Let us consider  the ratio of  the vacuum and the matter 
contributions to mixing:  
\begin{equation} 
R_{m/v}~\equiv~{ U_{e4} \sin 2 \psi \cdot
 \sqrt{2} G_F N_n^{res}  \over  
\displaystyle{\Delta  m^2_{21} \over E} \sin 2 \omega  }~.   
\label{off}
\end{equation}
Here the neutron density is taken in the resonance 
$N_n = N_n^{res}$  
which is justified for small $\omega$ (narrow resonance).  
Determining $N_n^{res}$  
from the resonance condition ($H_{22} = 0$, where $H_{22}$ is the 
22-element of the Hamiltonian 
(\ref{evolution})),  we find: 
\begin{equation} 
{R}_{m/v} =  
\frac{U_{e4}}{4 \tan 2\omega } 
\left[r_R \cos^2 \phi - {1\over {2}} (\sin^2 \psi
   - |U_{e4}|^2 \cos^2 \psi ) \right]^{-1} \approx 
\frac{U_{e4}}{4 \tan 2\omega } 
\left[r_R  - {1\over {4}} \right]^{-1}~,   
\label{off1}
\end{equation}
where the last equality holds for small 
$\phi$ and $\psi \approx \pi/4$. 
Here $r_R \equiv N_e/N_n$ is the ratio of the electron and  neutron  
densities in resonance. The ratio  increases from 2 in the center of the 
Sun to about 6 at the surface \cite{BP95}.   
Using the upper bound  $U_{e4} < 0.2$ 
from the CHOOZ experiment \cite{CHOOZ}  and the lower bound,   
${\rm tan}2 \omega > 0.06$,  implied by  a 
small mixing solution of the $\nu_{\odot}$-problem 
we get from (\ref{off1}) 
$R_{m/v} \leq 0.4$. That is, the matter effect is smaller than the
vacuum effect, although the difference is not large. The matter effect  
can both enhance and suppress the vacuum effect, thus enlarging 
a  range of possible vacuum mixing angles: 
$\sin^2 2 \omega = (0.5 - 3)\sin^2 2 \omega_0 $;  here 
$\omega_0$ is the angle without off-diagonal matter effect. 

If the  original boron neutrino flux is smaller than  the SSM one, 
then smaller total mixing is needed  and $R_{m/v}$ can be of the order
one. 
In this case the matter effect can give dominant contribution 
or even substitute the vacuum mixing in (\ref{evolution}).\\

For $U_{e4} \ll 0.2$ the matter term in 
the off-diagonal  element of $H_2$ is much smaller
than the vacuum term, and $P_{2MSW}$ reduces to usual 2$\nu$ 
survival probability characterized by  $\omega$, 
$\Delta m^2_{21}$ and   
$N^{eff} \approx \cos^2 \phi N_e - {1 \over 2 }\sin^2 \psi N_n $ 
(see  (\ref{solNeff})).   
Using the reactor bound on the admixture of $\nu_e$ in 
the state $\nu_3$:  
$\cos^2 \phi > 0.95$, we get  
$N^{sterile} < N^{eff} < N_e$, 
where $N^{sterile} = N_e - N_n/2$ is the effective density for pure
$\nu_e \rightarrow \nu_s$ 
conversion. As a consequence, in our case  the survival probability at
the adiabatic edge of 
the MSW suppression pit (survival probability as a function of 
$E$) is in between the probabilities for the 2$\nu$ sterile and 
2$\nu$ active cases; the  non-adiabatic edges 
(determined by the derivative 
$d[ln(N^{eff})]/dx$) practically coincide for all three cases.\\

Main feature of  solution of the solar neutrino problem in our scenario
which can be used
for its identification is that solar neutrino flux should contain
 both $\nu_{\mu}$ and $\nu_s$ neutrinos. 
Moreover, due to large
$\nu_{\mu}$ - $\nu_s$ mixing in the $\nu_2$-state implied by the
atmospheric neutrino data the transition probabilities $P(\nu_e
\rightarrow \nu_s)$ and $P(\nu_e \rightarrow \nu_{\mu})$ should be
comparable. (The $\nu_e$ $\rightarrow$ $\nu_{\tau}$
probability is small, since $\nu_{\tau}$ is weakly
mixed with other neutrinos.) We find: 
\begin{equation}
\begin{array}{c}
P(\nu_e \rightarrow \nu_{\mu}) \approx \cos^2 \phi  
\left[  \sin^2 \psi 
 \sin^2 \phi  P_{2MSW} + \cos^2 \psi  (1 - P_{2MSW}) \right .
~~~~~~~~~~~~~~~~~~~~~~~~~~\\
~~~~~~~~~~~~~~~\left . +  \sin {\phi}  \sin {2\psi}
\tan2 \omega ({1\over {2}} - P_{2MSW}) +  \sin^2 \phi  \sin^2{\psi}
\right]~~, 
\end{array}
\label {Pemu}
\end {equation}
\noindent
\begin{equation}
\begin{array}{c}
P(\nu_e \rightarrow \nu_s) \approx \cos {\phi}^2 
\left[ \cos^2 {\psi}
 \sin^2{\phi} P_{2MSW} \right.~~~~~~~~~~~~~~~~~~~~~~~~~~~~~~~
\\~~~~~~~~~~~~~~~\left. +  \sin^2 \psi (1 - P_{2MSW}) -  \sin {\phi}
\sin {2\psi}
\tan2 \omega ({1 \over {2}} - P_{2MSW}) +  \sin^2 \phi  \cos^2 \psi 
\right]~ .
\end{array}
\label {Pes}
\end {equation}
In the limit of small $\omega$ and small $\phi$  we get from 
(\ref{Pemu}, \ref{Pes}): 
\begin{equation} 
\begin{array}{c}
P(\nu_e \rightarrow \nu_{\mu}) \approx \cos^2{\psi} (1 - P_{2MSW})
\\
P(\nu_e \rightarrow \nu_s) \approx \sin^2{\psi} (1 - P_{2MSW})~. 
\end{array}
\end{equation}
That is, the relative contributions of these channels are determined
by the angle $\psi$. In particular, for $\sin^2 \psi \approx 0.5$ we
have  
$P(\nu_e \rightarrow \nu_{\mu}) \approx P(\nu_e \rightarrow
\nu_{s})$. Taking $\sin^2 2\psi \approx 0.70-1.0$ as it follows from
the atmospheric neutrino data we find
\begin{equation} 
$$\displaystyle { {P(\nu_e \rightarrow \nu_s) \over P(\nu_e \rightarrow
\nu_{\mu})}~=~\tan ^2 \psi~ =~{1 \over 4}~ -~4 }.$$
\end{equation} 
This ratio does not depend on the neutrino energy (fig. 3).

In what follows we will refer to this solution as to the mixed 
$\nu_e \rightarrow \nu_{\mu}, \nu_s$ solution. As we will see,  
its properties  are intermediate between 
properties of the pure flavor 
$\nu_e \rightarrow \nu_{\mu}$ 
and pure sterile 
$\nu_e \rightarrow \nu_s$ solutions.

Some general tests of presence of the sterile neutrinos in the 
solar neutrino flux have been elaborated in \cite{BandG2}. 
Transition to sterile neutrinos suppresses the number of neutral current
events. Let us consider a modification of the
charged-to-neutral current events ratio, 
$(CC/ NC)$,   
which will be measured in the SNO experiment.
Let us introduce the double ratio:
$$r_d \equiv \left({CC \over NC} \right)_{osc} : 
\left({CC \over NC}\right)_{0}~,
$$
where $({CC / NC})_{0}$ 
and $({CC / NC})_{osc}$ 
are  the ratios in absence of oscillations 
and with oscillations. For
three possible solutions of the $\nu_{\odot}$-problem we get:
\begin{equation} 
r_d= {\displaystyle \left\{ \begin{array}{cc}  <P_{ee}> &
~~~~~~~~~~~~~~~~\nu_e \rightarrow \nu_{\mu} \\ 
\displaystyle { { < {P}_{ee}>  \over {1-
< {P}_{es}> }}} & ~~~~~~~~~~~~~~~~~~\nu_e \rightarrow \nu_{\mu}, \nu_s 
\\
\sim 1 & ~~~~~~~~~~~~~~~~~~~ \nu_e \rightarrow \nu_s    
\end{array} \right . }
\label{douR1}
\end{equation} 
where $<...>$ means averaging over  energy spectrum of the boron
neutrinos. According to (\ref{douR1}):
\begin{equation} 
r_d(\nu_e \rightarrow \nu_{\mu} )~ < ~ r_d(\nu_e \rightarrow
\nu_{\mu}, \nu_s)~ < ~ r_d(\nu_e \rightarrow
\nu_s)~. 
\end{equation}
For small mixing solution and the SSM boron neutrino flux 
\cite{BP95}, 
we find intervals 
\begin{equation} 
r_d= {\displaystyle \left\{ \begin{array}{cc}  0.3~-~0.4 &
\nu_e \rightarrow \nu_{\mu} \\ 0.5~-~0.8 &~~ \nu_e
\rightarrow  \nu_{\mu}, \nu_s \\
1.07~-~1.09 & \nu_e \rightarrow \nu_s    
\end{array} \right . }~~~~ 
\label{rd}
\end{equation} 
which are well separated from each other. 
(In (\ref{rd}) difference of  energy dependences of the $CC$- 
 and $NC$- current cross-sections is taken into account, which leads to
$r_d > 1$ in the last line \cite{KrPe}.) 
Thus,  if the original flux of
the boron neutrinos, $F_B$,  
is known, it will be easy to disentangle solutions by
measuring $r_d$. However,  uncertainties in $F_B$ 
make the task to be ambiguous. If, {\it e.g.}, $F_B$ 
is smaller than the one in the SSM \cite{BP95}:  
$F_B \sim 0.7 F_B^{SSM}$, 
then for the ($\nu_e$ $\rightarrow$ $\nu_{\mu}$) solution we get $r_d \approx
0.4~-~0.6$ which overlaps with the interval expected for ($\nu_e$
$\rightarrow$ $\nu_{\mu}$, $\nu_s$).\\

Due to presence of  both $\nu_s$ and $\nu_{\mu}$ in the final state, 
the effect of the $NC$-interactions  in the $\nu e-$ scattering 
(the Super-Kamiokande experiment)  is 
 intermediate  between the effects in the pure active 
and pure sterile neutrino cases. 
Correspondingly,  the 
$\sin ^22 \omega - \Delta m^2_{21}$ 
region of solutions is intermediate between regions for 
pure active and sterile 
\cite{KLP} neutrino conversions. This gives   
$\sin ^2 2 \omega  = (4.5 - 11) \cdot 10^{-3}$ and 
$\Delta m^2_{21} =  (3  - 9) \cdot 10^{-6}$ eV$^2$ in the SSM framework.\\

Let us consider  the distortion of the energy spectrum of the
recoil electrons. In the SNO detector
$(\nu_e d \rightarrow e p p )$ 
the distortion is determined by the probability $P$($\nu_e$
$\rightarrow$ $\nu_e$)$\approx \cos^4 \phi \cdot P_{2MSW}$. 
The factor
$\cos^4 \phi$ leads to smoothing of energy dependence of the
probability, and therefore the distortion of the recoil electron
spectrum can  be weaker than in the pure 
$\nu_e \rightarrow \nu_{\mu}$ 
or $\nu_e \rightarrow \nu_s$ case. 
However for $\sin^2 2\phi <
0.2$ this effect is  small ($\cos^4 \phi \sim 0.95$). The 
Super-Kamiokande experiment $(\nu e \rightarrow \nu e )$ is sensitive to
both 
$CC$- and $NC$- interactions. The
$NC$-scattering of $\nu_{\mu}$ leads to smoothing of the
spectrum distortion. The  transition to sterile neutrinos removes
this effect. Therefore in the case of mixed conversion,  
$\nu_e \rightarrow \nu_{\mu}, \nu_s$, the spectrum is 
distorted stronger than in the
$\nu_e$ $\rightarrow$ $\nu_{\mu}$ 
case but weaker than due to $\nu_e$ $\rightarrow$ $\nu_s$ 
conversion  \cite{BLS} (fig. 4),  
although  the difference is  small. 
\\

Let us discuss  the day-night effect \cite{DN}.
The $\nu_e \rightarrow \nu_s,  \nu_{\mu}$ gives  
 weaker asymmetry than $\nu_e \rightarrow \nu_{\mu}$,    
but stronger asymmetry than $\nu_e \rightarrow \nu_s$.     
The last two cases  were studied in  \cite{KLP} and \cite{MPet}.   

An  important signature of our scenario is the day-night effect for
events induced by the neutral currents, $N^{NC}$. 
(This in pure $\nu_e \rightarrow \nu_s$ conversion case 
 has been studied in \cite{KLP}.) 
Let us introduce the
day-night asymmetry as
$$
A^{NC}_{D/N}~=~{ {N^{NC}_{N}~-~N^{NC}_{D}} \over
{N^{NC}_{N}~+~N^{NC}_{D}}}~.  
$$
(Similarly,  the  asymmetry for $CC$-events, $A^{CC}_{D/N}$,  
can be defined).  
The asymmetry can  be
used to distinguish our scenario from pure 2$\nu$-cases. 
Indeed,  $A^{NC}_{D/N}$ is zero for the 
$\nu_e \rightarrow \nu_{\mu}$ 
conversion and non-zero in presence of sterile neutrinos. 
The intermediate  case ($\nu_e$ $\rightarrow \nu_{\mu}, \nu_s$) can be
distinguished from the pure sterile case by comparison of asymmetries in
the charged currents, $A^{CC}_{D/N}$,  and in the neutral currents. For
pure sterile case: $A^{CC}_{D/N}~-~A^{NC}_{D/N} \approx 0$  \cite{KLP}.  
In the ($\nu_e \rightarrow \nu_{\mu}, \nu_s$) case 
we predict $A^{CC}_{D/N}~ - ~A^{NC}_{D/N} >  0$. Studies of
these asymmetries will be possible in the SNO experiment.


\vskip 0.5cm
\leftline{\bf 5. Short range oscillation experiments. LSND.}
\vskip 0.2cm

In our scenario, two neutrino states with large masses, $\nu_3$ and
$\nu_4$, can be relevant for short range experiments. 
The probability
of $\nu_{\mu} \leftrightarrow \nu_e$ 
($\bar{\nu}_{\mu} \leftrightarrow \bar{\nu}_e$) 
oscillations can be written as
\begin{equation}
\begin{array}{c} 
P(\nu_{\mu} \rightarrow \nu_e) \approx 
4 |U_{\mu3}|^2|U_{e3}|^2 
\sin^2 \left( {\displaystyle m_3^2 t \over {4E}} \right)
+ 4 |U_{\mu4}|^2|U_{e4}|^2 
\sin^2 \left({\displaystyle m_4^2 t \over {4E}} \right)
~+ ~~~~~~~~~~~~~~~~~~

\\
4Re[U^*_{e3} U_{\mu3} U_{e4}
U^*_{\mu4}] \cdot \left\{ \sin^2 \left [ \displaystyle {
{(m^2_4-m^2_3)t} \over {4E}} + 
\displaystyle { \delta \over 2} \right] - 
\sin^2 \left(\displaystyle{{m^2_4 t} \over {4E}} +
\displaystyle { \delta \over 2}\right)
- \sin^2 \left (\displaystyle{{m^2_3 t} \over {4E}} - 
\displaystyle { \delta \over 2}\right)
\right \}~, 
\end{array}
\label{PLSND}
\end{equation} 
where $\delta=arg(U^*_{e3} U_{\mu3} U_{e4} U^*_{\mu4})$. For real
mixing matrix we get $\delta = 0$. The first two terms in (\ref{PLSND})
correspond to ``indirect'' oscillations \cite{indirect} 
related to levels $\nu_3$ and
$\nu_4$ correspondingly. The third term is the result of interference
of oscillations due to $\nu_3$ and $\nu_4$. 
For $m_3 \sim O({\rm eV})$   
we get the following results
depending on value of the mass $m_4$:  

1) The mass  $m_4 \le 0.3$ eV  
is too small to contribute to the LSND effect 
and the problem  is reduced to the one level problem  with
probability described by the first term in (\ref{PLSND}). 
For values of $U_{\mu3}$ and  $U_{e3}$ 
similar to those in  quark sector we  get 
negligibly small probability for LSND: $P\le10^{-6}$.  
If  $U_{\mu3}$ and $U_{e3}$ are at
the upper  experimental bounds, then  $P\le10^{-3}$ which
could  be reconciled  with data at 3$\sigma$-level \cite{BandG}.
For $m_4 \leq 0.1$ eV, one can reproduce a desired zenith angle 
dependence in atmospheric neutrinos.

2) For  $m_4 \geq 0.3$ eV   both
$\nu_3$ and $\nu_4$ can give sizable effect to  the 
short range oscillations. Also
interference between two different modes of oscillations becomes
important. In certain ranges of parameters the interference is
constructive,  thus leading to substantial enhancement of the
probability. For instance,  if  $m^2_3 =  1$ eV$^2$ and $m^2_4 =  0.5$
eV$^2$, the probability 
$P$($\nu_{\mu}$ $\rightarrow$ $\nu_e$) is enhanced by a factor
 of more than 4,  as compared  to the  one level case. Now the 
probability  $P\sim 2.2\cdot 10^{-3}$ turns out to be in a 
good agreement with LSND result. 
In this case one also predicts $\nu_e$ $\leftrightarrow$ $\nu_x$  and
$\nu_{\mu}$ $\leftrightarrow$ $\nu_x$ oscillations at the level of
present experimental bounds. 
However, no zenith
angle dependence of atmospheric neutrino deficit  is expected.\\ 


\vskip 0.5cm
\leftline{\bf 6. Supernova neutrinos}
\vskip 0.2cm

Since $\nu_{\mu}$ and $\nu_s$ are (almost)
maximally mixed, their level crossing occurs at zero (small) density. 
Let us
introduce the eigenstates of system ($\nu_{\mu}$,  $\nu_s$) in medium
$\nu'_{2m}$, $\nu'_{4m}$, in absence of  mixing with $\nu_e$ and 
$\nu_{\tau}$. At zero density $\nu'_{2m} \approx \nu_2$ and  $\nu'_{4m}
\approx \nu_4$. Using the eigenvalues of 
$\nu'_{2m}$ and $\nu'_{4m}$ 
it is easy to find that 
whole system ($\nu_e$ and $\nu_{\tau}$ included) 
has four resonances (level crossings) in the 
neutrino channels (fig. 5):  
(i) (${\nu}_{\tau}$ - ${\nu}'_{4m}$) 
crossing at density $N_{\tau 4}$,  
(ii) ($\nu_e$ - $\nu_{\tau}$) crossing  at density 
$N_{e\tau}$, 
(iii) ($\nu_e$ - $\nu'_{4m}$) crossing  at $N_{e4}$, and  
(iv) ($\nu_e$ - $\nu'_{2m}$) crossing  at $N_{e2}$.        
The mass hierarchy leads to hierarchy of the resonance densities (see
fig. 5):
$$
N_{\tau 4} \geq N_{e \tau} \gg N_{e 4} \geq N_{e 2}. 
$$
There is  no  resonances in the antineutrino channels.  
For $m_3 \le 5$ eV we get  $m_N N_{\tau 4} \le 10^8~ {\rm g/cm}^3$
(here $m_N$ is the nucleon mass). 
That is, the 
${\nu}_{\tau} \rightarrow {\nu}_s$ resonance conversion 
occurs outside the core. Moreover, the effective 
${\nu}_{\tau} - {\nu}_s$ mixing in the core is very small: 
$\sin^2 2\theta^m_{\tau 4} ~\approx~4|U_{s3}|^2 
\cdot N_{\tau 4}/N_{core}~, $ and 
generation of the sterile neutrinos is strongly suppressed.  
Consequently, there is no influence of the sterile
neutrinos on the gravitational collapse and cooling of the core.

The nucleosynthesis of heavy elements due to the 
$r$-processes in the inner parts of supernovas 
implies that $\nu_{\tau}$ $\rightarrow$ $\nu_e$ 
conversion does not occur efficiently above certain densities \cite{Qian}, 
so that there is no  electron neutrinos 
with hard spectrum in the range ($10^5 - 10^{10}$) g/cm$^3$. 
The mass  spectrum under consideration realises the possibility 
suggested in \cite{pelt} which allows one to  enhance 
the $r$-processes. Indeed, suppose that   transitions 
$\nu_{\tau} \rightarrow \nu_s$ 
and $\nu_e \rightarrow \nu_{\tau}$  
due to  both high density resonances   are efficient. 
In this case in the region of the 
$r$-processes the $\nu_e$ - flux is absent or strongly suppressed 
(see fig. 5), thus 
creating neutron rich medium needed for production of 
the heavy elements. \\

Let us consider consequences of  transitions for 
neutrino fluxes detectable  at the earth. 
Suppose first that all resonances are 
effective. In this case the following transitions occur:  
\begin{equation} 
\begin{array}{l}
\nu_e \rightarrow \nu_{\tau}~,~~~~ 
\nu_{\mu} \rightarrow \nu_e~ ,~~~~
\nu_{\tau}\rightarrow \nu_4 \rightarrow \nu_e~  
\rightarrow \nu_2~ (\approx \nu_s, \nu_{\mu}).\\ 
\end{array} 
\label{SNtr}
\end{equation}
Thus one expects (i)  that  
$\nu_e$-flux will have hard spectrum corresponding to 
the original $\nu_{\mu}$ - spectrum; (ii) in contrast, 
$\nu_{\tau}$'s will have soft spectrum (corresponding to the original
$\nu_e$- spectrum); (iii) the flux of
$\nu_s$'s is produced, of the order of $\sim 1/12$ of the total flux.  

Suppose now that low density resonances 
 $\nu_e - \nu_{2m}$ and  $\nu_e - \nu_{4m}$ are unefficient
(which may occur in compact stars with small mass and 
large density gradient in the outer layers). Then transitions proceed
as follows:
\begin{equation}
\nu_e \rightarrow \nu_{\tau}~; ~~~~  
\nu_{\mu} \rightarrow \nu_2~(\approx \nu_{\mu}, \nu_s); ~~~~ 
\nu_{\tau} \rightarrow \nu_e \rightarrow \nu_4~ 
(\approx \nu_{\mu}, \nu_s). 
\end{equation} 
Thus $\nu_e$'s  disappears  which can be established in 
experiment, $\nu_{\tau}$'s will have soft spectrum of the original
$\nu_e$, the  $\nu_s$-flux will  compose $\sim 1/6$ of the total flux. 
There is no transitions in the antineutrino channels.

Finally,  suppose that $\nu_e$ $\rightarrow$ $\nu_{\tau}$ resonance is
not efficient because of smallness of $U_{e3}$. 
The following transitions take place:
\begin{equation}
\nu_e \rightarrow \nu_4 ~ (\approx \nu_{\mu}, \nu_s)~;~~~~
\nu_{\mu} \rightarrow \nu_e ;~~~~
\nu_{\tau} \rightarrow \nu_2 ~ (\approx \nu_{\mu}, \nu_s).  
\end{equation}
In this case the $\nu_e$-flux in the  $r$-processes region 
is unchanged. At the detectors, however, the 
the $\nu_e$-flux will have hard spectrum. 
The $\nu_{\mu}$- and  $\nu_s$- fluxes will have both a hard component  
due to  the $\nu_{\mu} \rightarrow \nu_s$   convertion 
and a soft component due to the $\nu_e$ $\rightarrow$ $\nu_4$
convertion. 


\vskip 0.5cm
\leftline{\bf 7.  Primordial Nucleosynthesis }
\vskip 0.2cm
The oscillations $\nu_{\mu}$ $\leftrightarrow$ $\nu_s$, $\bar{\nu}_{\mu}$
$\leftrightarrow$ $\bar{\nu}_s$ with parameters implied by the atmospheric
neutrino problem result in appearance of the equilibrium concentration 
of sterile neutrinos in the early Universe.  
Therefore, during  the time of nucleosynthesis, $t \geq 1 s$, the
effective number of neutrino species is $N_{\nu} = 4$. This 
disagrees with bound $N_{\nu} < 2.6$ based on low values of
deuterium abundance \cite{cris}. Four neutrino species are, 
however, in agreement with recent 
conservative estimations  $N_{\nu} < 4.5$ \cite{sar,ncris} which use 
large abundance of $^4He$. In view of
these uncertainties let us comment on possibility to suppress 
generation of  sterile neutrinos. 
It was shown  that leptonic asymmetry
in active neutrinos $L_{\nu} \equiv (n_{\nu}-n_{\bar{\nu}})/n_{\gamma} >
7 \cdot 10^{-5}$ at the time $t > 10^{-1} s$  \cite{Foot}, 
is enough to 
suppress $\nu_{\mu} \leftrightarrow \nu_s$ oscillations. 
Such a $L_{\nu}$ produces the potential due to $\nu - \nu$
scattering:  $V \sim G_F L_{\nu} n_{\gamma}$ which suppresses the
effective mixing angle in matter:
$\sin^2 2\theta_m ~\sim~\sin^2 2\theta \Delta m^2/ T V$ 
($T$ is the temperature).  
Although there is no strong restrictions on leptonic asymmetry one
would expect that $L_{\nu}$ is of the order of the baryon asymmetry:
$L_{\nu} \sim 10^{-10}$. 
It is argued \cite{Foot2} (see also \cite{shi}) that 
large $L_{\nu}$ can be produced
in oscillations of the tau neutrinos to sterile
neutrinos: $\nu_{\tau} \rightarrow \nu_s$, 
$\bar{\nu}_{\tau} \rightarrow \bar{\nu}_s$ 
in earlier epoch ($t < 10^{-2}$ s). 
In presence of small original leptonic asymmetry, $L_{\nu} \sim
10^{-10}$, these oscillations can lead to exponentially fast increase of
$L_{\nu}$ (the task is non-linear). For $m_3 \geq 1$ eV, the increase
occurs at $T \sim 16$ MeV. 
The lower value of $m_3^2$ required by this mechanism is approximately
proportional to  $\Delta m_{42}^2$. Thus,  
for $\Delta m_{42}^2 = 3 \cdot 10^{-3}$ eV$^2$ 
we get from \cite{Foot2} $m_3^2 \geq 6$ eV$^2$ and 
$\sin^22\theta_{\tau s} \leq 8 \cdot 10^{-6}$.   
For $\Delta m_{42}^2 = 10^{-3}$ eV$^2$  
the corresponding bounds are 
$m_3^2 \geq 2$ eV$^2$  and 
$\sin^22\theta_{\tau s} \leq 4 \cdot 10^{- 5}$.   
For larger angles the $\nu_{\tau}$ $\rightarrow$ $\nu_s$
oscillations themselves produce the equilibrium concentration of sterile
neutrinos. These conditions can be satisfied in our scenario.

\newpage

\vskip 0.3cm
\leftline {\bf 8. Properties of mass matrix}
\vskip 0.2cm 

Let us find the elements of the neutrino mass matrix in the flavor basis, 
$||m_{\alpha \beta}||$, $(\alpha, \beta=e,\mu,\tau, s)$
which lead to the  scenario under discussion (fig. 1). 
Suppose first that 
mixing of $\nu_{\tau}$ is very small and it does not influence masses
and mixing of lighter neutrinos. Then in the rest system:  
($\nu_s$, $\nu_e$,  $\nu_{\mu}$) the  
sub-system ($\nu_s, \nu_{\mu}$) with bigger masses and
approximately maximal mixing dominates. This means that we can find masses
and mixing of $\nu_s$ and $\nu_{\mu}$ considering the 
$2 \times 2$ sub-matrix   
$||m^{(2)}_{\alpha \beta}||$, with  $\alpha, \beta = s,\mu$ only.
The effect of mixing with $\nu_e$ gives small corrections. 

Main feature of our scenario is the maximal mixing 
of two states with at least one order of magnitude mass hierarchy:  
$\epsilon \equiv m_2/m_4 = 0.03 - 0.10$. 
Maximal mixing and strong 
hierarchy imply that sub-matrix $||m^{(2)}_{\alpha \beta}||$ 
has  small determinant and all its elements are of the same order. 
Indeed, we find  the following relations: 
$$
\displaystyle { {m_{ss}~-~m_{\mu \mu} \over m_{\mu s}} ~=~ {2 \over
{\rm tan}2\theta_{atm}}}, 
$$  
$$
{m_{\mu \mu}m_{ss}~-~m^2_{\mu s} \over 
({m_{\mu \mu}~+~m_{ss}})^2 } ~=~ 
\frac{\epsilon}{(1 + \epsilon)^2}
$$ 
and $m_4 \approx m_{\mu \mu}~+~m_{ss}$. 
For maximal mixing the masses are determined by  
\begin{equation} 
m_{ss}~=~m_{\mu \mu}~\approx~m_{\mu s} \cdot 
\frac{1 + \epsilon}{1 - \epsilon}~.
\label{mssmuu}
\end{equation} 
For $\ 0.03 - 0.1$,  the eq. (\ref{mssmuu}) gives 
$m_{ss} \approx (1.05 - 1.2) \cdot m_{\mu s}$ 
or 
$m_{ss} \approx - (0.95  - 0.8) \cdot m_{\mu s}$. 
In the case of non-maximal mixing 
larger spread of masses is possible. For $\sin^22\theta_{atm} = 0.9$  we
find: $m_{ss} = m_{\mu \mu} + m_{\mu s}$,  and 
$m_{\mu \mu} ~:~ m_{\mu s}~:~ m_{s s}$ = $0.67~:~1~:~1.3$. 
Thus,  a spread of the elements can be within factor of 2. 
Taking  $m_4 \approx 2 m_{\mu s}$,  we get: 
$m_{\mu s} \sim (1 - 3)\cdot 10^{-2}$ eV. 

Diagonalizing the sub-matrix $m^{(2)}$ 
one  can find the mass
matrix for the light neutrino system ($\nu_e$,  $\nu'_2$), where $\nu'_2$
is the  neutrino with mass $m_2$.  From this we get  the mixing 
of the light neutrinos:  
\begin{equation} 
\sin \theta_{\odot}~\approx~\displaystyle{1\over m_2} 
(m_{e \mu }\cos \theta_{atm}~-~ m_{es}\sin \theta_{atm}) 
~\approx~ 
(3~-~5) \cdot 10^{-2}~,  
\label{mee}
\end{equation} 
and if there is no strong cancellation in (\ref{mee}):   
$$
m_{e \mu},~m_{e s}~\le~\sqrt{2} \sin \theta_{\odot} \cdot m_2
~\approx~ 2\cdot 10^{-4} ~{\rm eV}~.
$$

The structure of the mass matrix can be substantially different if 
the mixing of the tau neutrino is not small. In particular, 
the desired mass $m_{ss}$ can be generated by this mixing: $m_{ss} 
\approx -(m_{s\tau})^2 / m_{\tau \tau}$, 
if, {\it e.g.}, $m_{s \tau} \sim 0.1$ eV  and $m_{\tau
\tau} \approx (1-2)$ eV. This corresponds to rather large 
$\nu_{\tau} - \nu_s$ mixing parameter:  
$\sin^22\theta_{\tau s} \sim (2-4) \cdot 10 ^{-2}$ for which  
the equilibrium concentration of $\nu_s$ is
produced in the early
Universe already in $\nu_{\tau} \leftrightarrow \nu_s$
oscillations. The mechanism of generation of the lepton 
asymmetry \cite{Foot} does not work. 

As follows from the above estimations of  the mass terms, the mass
matrix for the active
neutrinos can have usual hierarchical structure with small
mixing. It can be easily generated by the see-saw mechanism 
with linear hierarchy of masses of the right handed neutrinos.  
Also  $\nu_s$ can  have the hierarchy of couplings with different
generations: $m_{es} \ll m_{\mu s} \ll m_{\tau s}$. 
Some elements can be zero: {\it e.g.}, $m_{es} =  m_{\tau s} = 0$. 


\vskip 0.3cm 
\noindent
{\bf 9. Conclusion} 
\vskip 0.2cm

We have considered minimal modification of the 
``standard'' scenario for the neutrino mass introducing one 
additional sterile
neutrino. The sterile neutrino mixes strongly with muon neutrino, so that
the $\nu_{\mu} \leftrightarrow \nu_s$ oscillations solve the atmospheric
neutrino problem. 

We show that the parametric enhancement 
of the $\nu_{\mu} \leftrightarrow \nu_s$ oscillations occurs 
when  high energy atmospheric neutrinos cross the core of the Earth. 
This effect can be relevant for explanation of the anomaly 
in the zenith angle distribution observed by the  MACRO experiment. 
Suppression of the $NC$-events in the 
atmospheric neutrinos ($\pi^0$ - like events) is 
another signature of the scenario.

The scenario can supply the hot dark matter of the Universe.  

The solar neutrino problem is solved by  $\nu_e$
conversion to $\nu_{\mu}$ and $\nu_s$. The solution has 
characteristics  (spectra distortion, day-night effect, $NC/CC$ ratio) 
being intermediate between characteristics of $\nu_{e}\rightarrow \nu_{\mu}$
and  $\nu_{e} \rightarrow \nu_s$ conversions.

The probability of 
$\bar{\nu}_{\mu} \rightarrow \bar{\nu}_e$ 
oscillations in the LSND experiment can reach
O($10^{-3}$). It can be further enhanced if new neutrino state has the
mass $m_4 \geq 0.3$ eV. (In this case, however, the atmospheric neutrino
deficit will have no zenith angle dependence.)

Disappearance of the $\nu_e$-flux from supernovas 
is another possible signature. 

The scenario can be
well identified or rejected by the neutrino experiments 
of new generation (Super-Kamiokande,
SNO, Borexino, ICARUS and others ).

\vskip 0.3cm
\leftline{\bf Acknowledgments.} 

The authors are grateful to E. Akhmedov and S. P. Mikheyev for useful
discussions, and to R. Volkas and X. Shi for valuable comments. 
This work has been accomplished  during 
the Extended workshop  on Highlights in Astroparticle Physics
(ICTP,  Trieste, October - December,  1997). The work of Q.Y.L. is
supported in part by the EEC grant ERBFMRXCT960090.

\newpage 

\noindent
{\bf Figure Captions} \\

\noindent
Fig. 1. Qualitative pattern of the neutrino masses and mixing. 
Boxes correspond to different mass eigenstates. The sizes 
of different regions in the boxes determine flavors 
of  mass eigenstates, $|U_{\alpha i}|^2$. 
White regions correspond to  the sterile flavor, 
light and dense shadowed regions fix the electron and muon 
flavors correspondingly; admixtures of the tau  
flavor are shown in black.\\

\noindent
Fig. 2. The zenith angle dependence of the survival probability 
$P(\nu_{\mu} - \nu_{\mu})$ for  atmospheric neutrinos for different 
values of the neutrino energy (figures at the curves in GeV) 
and $\Delta m^2 = 5 \cdot 10^{-3}$ eV$^2$. Solid 
lines correspond to 
$\nu_{\mu} \leftrightarrow \nu_s$ oscillations, 
dashed line is for   
$\nu_{\mu} \leftrightarrow \nu_{\tau}$ oscillations.\\

\noindent
Fig. 3.  The conversion probabilities of solar neutrinos 
as functions of $E/\Delta m^2$ for $\sin^2 2\omega = 0.01$, 
$\cos^2 \psi = 0.5$ and $\sin \phi = -0.2$;  
$P(\nu_e \rightarrow \nu_e)$ is shown by solid line,  
$P(\nu_e \rightarrow \nu_{\mu})$ -- dotted line,   
$P(\nu_e \rightarrow \nu_s)$ -- dashed line.\\

\noindent
Fig. 4. The expected distortion of the recoil electron energy spectrum 
in the Super-Kamiokande experiment. The ratio 
of numbers of events with and without conversion 
$R_e$ is shown by histograms:  bold solid line corresponds to  
$\nu_{e} \rightarrow \nu_s, \nu_{\mu}$ conversion 
considered in this paper;   
solid line is  for  
$\nu_{e} \rightarrow \nu_s$ conversion, and 
dotted line is for $\nu_{e} \rightarrow \nu_{\mu}$ conversion. 
The following values of parameters were used: 
$\Delta m^2 = 5 \cdot 10^{-6}$ eV$^2$, 
$\sin^2 2\omega = 8.8 \cdot 10^{-3}$, 
and (for our scenario) 
$\cos^2 \psi = 0.5$ and $\sin \phi = 0$. 
Also the Super-Kamiokande experimental points  from 306 days of observation 
are shown (statistical errors only). \\

\noindent
Fig. 5.  The level crossing scheme for supernova neutrinos. 
Solid lines show  the eigenvalues of 
 $\nu_{3m}$, $\nu_{4m}$, $\nu_{2m}$ and $\nu_{1m}$. 
Black dots  represent 4 resonances.  
Dashed lines correspond to energies of $\nu_e$, $\nu_{\tau}$ (straight
lines) and  $\nu'_{4m}$, $\nu'_{2m}$.

\newpage

\begin{figure}[H]
\mbox{\epsfig{figure=Fig1.eps,width=15cm,height=15cm,angle=0}}
\vglue1.0cm
\caption[]{ }
\end{figure}

\begin{figure}[H]
\mbox{\epsfig{figure=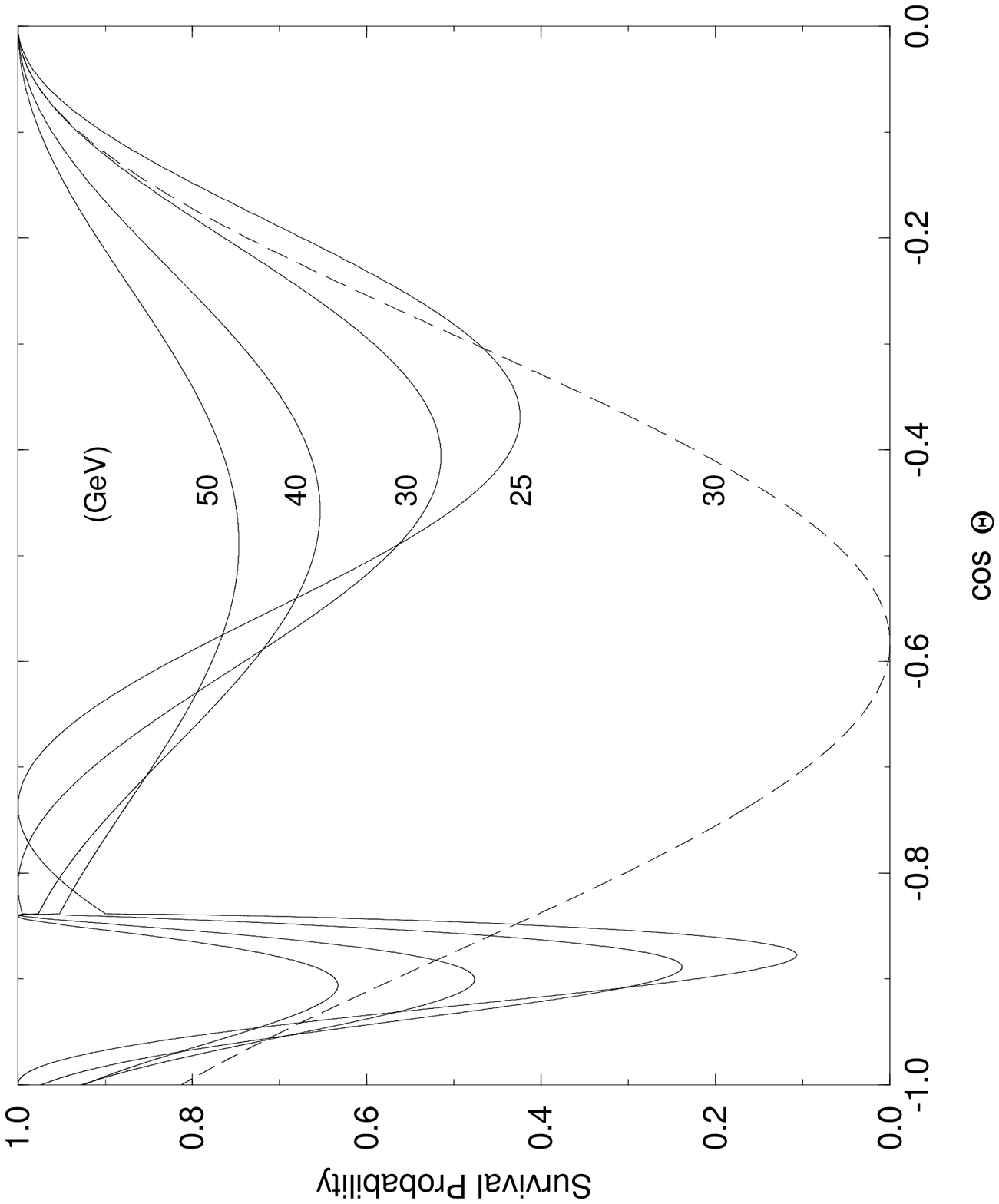,width=15cm,height=20cm,angle=0}}
\vglue1.0cm
\caption[]{ }
\end{figure}

\begin{figure}[H]
\hglue1.5cm 
\mbox{ \epsfig{figure=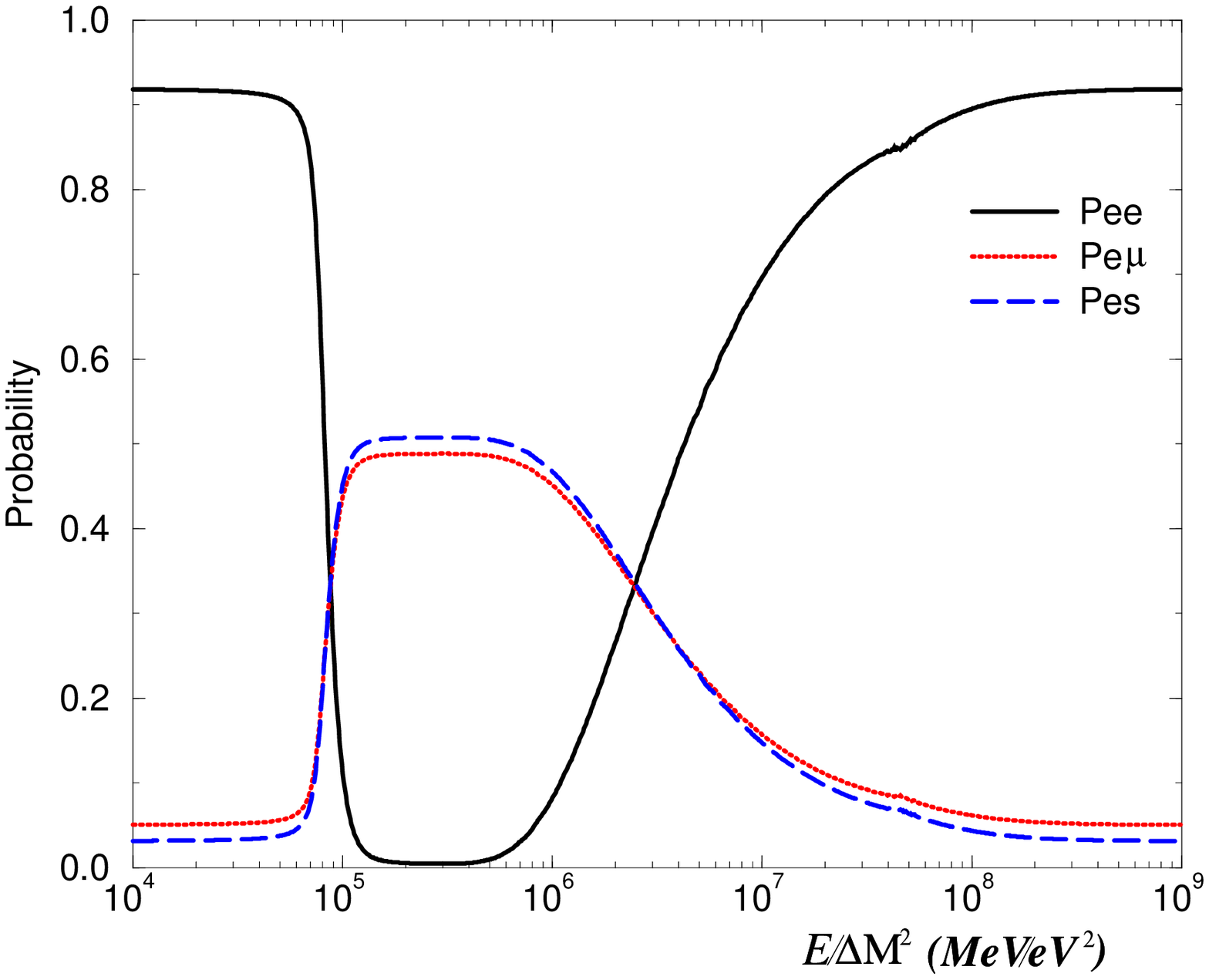,width=15cm,height=20cm,angle=0}}
\vglue-0.1cm
\caption[]{ }
\end{figure}

\begin{figure}[H]
\mbox{ \epsfig{figure=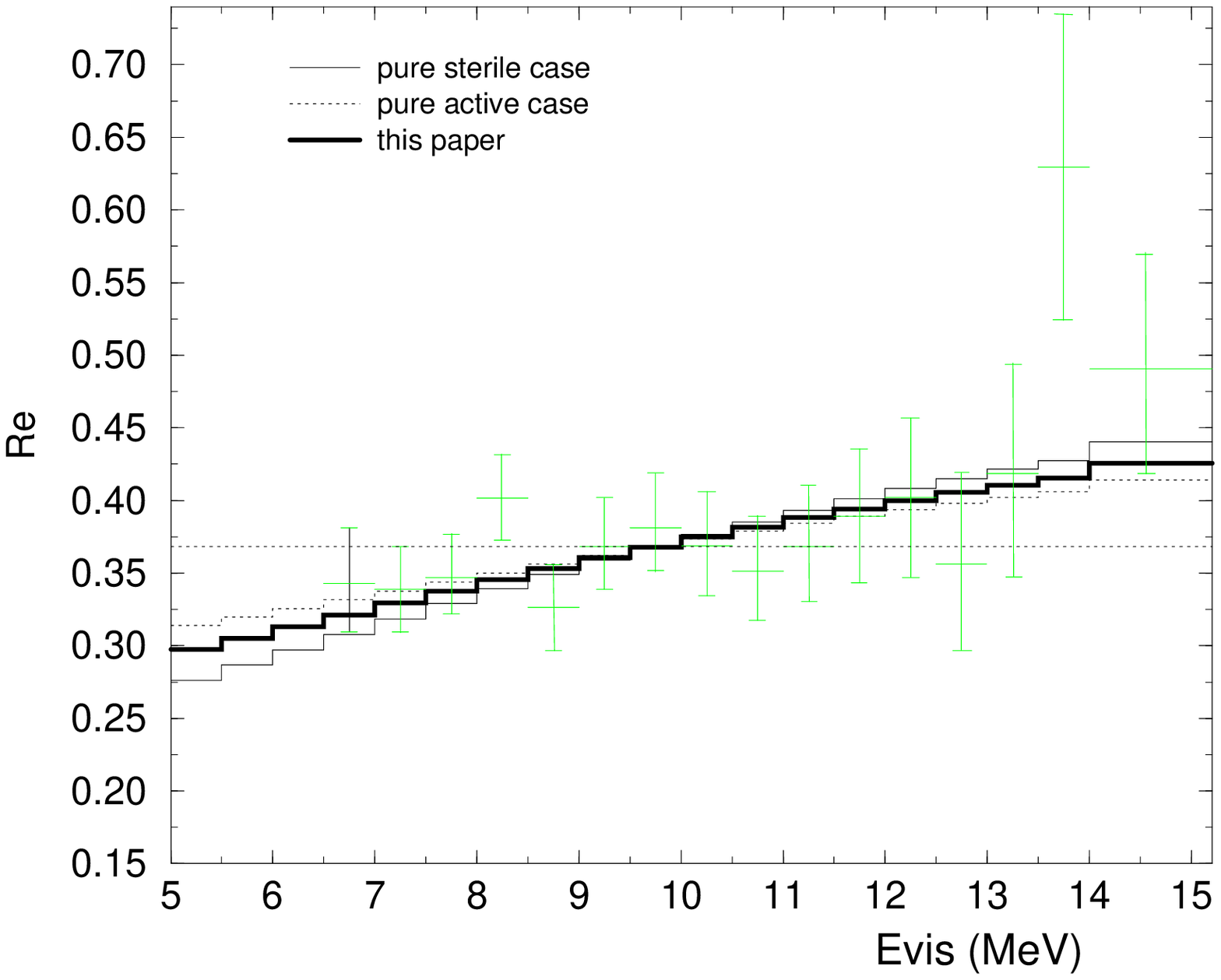,width=15cm,height=20cm}}     
\vglue0.5cm
\caption[]{ }
\end{figure}

\begin{figure}[H]
\vglue 0.7cm
\mbox{ \epsfig{figure=Fig5.eps,width=15cm,height=18cm,angle=0}}     
\vglue0.5cm
\caption[]{ }
\end{figure}


\begin{thebibliography}{99}

\bibitem{Alexei1} See for review A. Yu. Smirnov, 
Proc. of the  28th International Conference on High-energy Physics
(ICHEP 96), Warsaw, Poland, 25-31 July 1996, p. 288 (hep-ph/9611465). 

\bibitem{SW1} M. Gell-Mann, P. Ramond and R. Slansky, in: $Supergravity$, eds.
F. van Nieuwenhuizen and D. Freedman (Amsterdam, North Holland, 1979)
p. 315; \\
T. Yanagida, in: {\it Workshop on the Unified Theory and Baryon Number in the
Universe}, eds. O. Sawada and A. Sugamoto (KEK, Tsukuba) 95 (1979).

\bibitem{SW2} R.N. Mohapatra and G. Senjanovic, Phys. Rev. Lett. {\bf 44}
 (1980) 912.

\bibitem{Alexei2} See {\it e.g.},  A.Yu. Smirnov,
Nucl. Phys.  {\bf B466} (1996), 25.

\bibitem{indirect} A. De Rujula et al., Nucl. Phys.  {\bf B 168}
(1980), 54; V. Barger and K. Whisnant, Phys. Lett. {\bf B 209} (1988)
365;
S. M. Bilenky et al., Phys. Lett. {\bf B 276} (1992) 223; 
H. Minakata, Phys. Lett. {\bf B 356} (1995) 61;
K.S. Babu, J. Pati, F. Wilczek, Phys. Lett. {\bf
B 359} (1995) 351.

\bibitem{lsnd} C. Athanassopoulos et al, (LSND Collaboration)  
Phys. Rev. Lett., {\bf 77} (1996) 3082;  nucl-ex/9706006; nucl-ex/9709006. 

\bibitem{BandG} S. M. Bilenky, C. Giunti, W. Grimus, hep-ph/9711311.  


\bibitem{atm} Y. Fukuda et al.,  Phys. Lett. {\bf B 335} (1994) 237; 
D. Casper et al., Phys. Rev. Lett., {\bf 66} (1991) 2561; R. Becker-Szendy 
et al., Phys. Rev. {\bf D 46}, (1992) 3720; Nucl. Phys. B (Proc. Suppl.)
{\bf 38} (1995) 331. 

\bibitem{SK} Y.Totsuka (Super-Kamiokande Collaboration), in $LP'97$,
28th Int. Symposium on Lepton Photon Interactions, Hamburg,
Germany, 1997, to appear in the Proceedings;\\
E. Kearns, talk given at conference on $Solar~Neutrinos:~News~About~SNUs$,
December 2-6, 1997, Santa Babara.

\bibitem{Soudan} S.M. Kasahara et al., 
Physical Review {\bf D55} (1997) 5282; T. Kafka, talk given at Fifth
Int. Workshop TAUP-97, September 7 - 11, 1997, Gran Sasso, Italy.

\bibitem{Fuller} C.Y. Cardall and G.M. Fuller, Phys. Rev. {\bf
D 53} (1996) 4421; G.L. Fogli, E. Lisi, D. Montanino, G. Scioscia,
Phys. Rev. {\bf D 56} (1997) 4365.     

\bibitem{cald} {D. O. Caldwell and R. N. Mohapatra},
Phys. Rev. {\bf  D 48} (1993) 3259;
{S. T. Petcov and A. Yu. Smirnov},
Phys. Lett., {\bf B 322} (1994) 109.
{A. Joshipura}, Z. Phys. {\bf C 64}  (1994) 31;
{ D. O. Caldwell and R. N. Mohapatra}, Phys. Rev. {\bf D 50}  (1994)
3477;
{ A. Ioannissyan and J. W. F. Valle}, Phys. Lett. {\bf  B 332}
(1994) 93;
{ B. Bamert and C. P. Burgess}, Phys. Lett. {\bf  B 329} (1994) 289;
{ D. G. Lee and R. N. Mohapatra}, Phys. Lett. {\bf B 329} (1994) 463;
{A. S. Joshipura},  Phys. Rev. {\bf  D 51} (1995) 1321; Hisakazu
Minakata, Osamu Yasuda, hep-ph/9712291; Phys. Rev. {\bf D 56} (1997) 1692.  


\bibitem{Akhmedov} E. Akhmedov, P. Lipari, M. Lusignoli, 
Phys. Lett. {\bf B 300} (1993) 128. 

\bibitem{FootandVol} R. Foot, R.R. Volkas, O. Yasuda, TMUP-HEL-9710, 
 (hep-ph/9710403); TMUP-HEL-9707, Sep 1997,
(hep-ph/9709483);\\
J. Bunn, R. Foot, R.R. Volkas, Phys. Lett. {\bf B 413} (1997) 109; \\
R. Foot, R.R. Volkas, Phys. Rev. {\bf D 52} (1995) 6595. 

\bibitem{CHOOZ} M. Appolonio et al., CHOOZ collaborator,
hep-ex/9711002.

\bibitem{LisiAlexei} E. Lisi and A. Yu. Smirnov, in preparation. 


\bibitem{earth} See {\it e.g.} F. D. Stacey, 
{\it Physics of the Earth} (John Wiley and Sons, New York, 
$2^{nd}$ edition, 1977. 


\bibitem{param} V.K. Ermilova, V.A. Tsarev and V.A. Chechin, Kr. Soob,
Fiz.Lebedev Institute 5 (1986) 26; E. Akhmedov, Yad. Fiz. 47 (1988) 475. 

\bibitem{param2} P.I. Krastev and A.Yu. Smirnov, Phys. Lett. {\bf
B 226} (1989) 341. 

\bibitem{MACRO} F. Ronga, Proceedings of the 17th International
Conference on Neutrino Physics and Astrophysics, page 529, edited by
K. Enqvist, K. Huitu and J. Maalampi. 

\bibitem{BAKSAN} M.M. Boliev, et al., Eighth Rencontres de Blois,  
$Neutrinos,~dark~matter~and~the~universe$, edited by T. Stolarcyk,
J. Tran Thanh Van, F. Vannucci, (1997) 296. 


\bibitem{LMS} Q.Y. Liu, S.P. Mikheyev and A.Yu. Smirnov, IC/98/30,
hep-ph/9803415. 

\bibitem{K2K} Y. Suzuki, Proceedings of the 17th International
Conference on Neutrino Physics and Astrophysics, page 237, edited by
K. Enqvist, K. Huitu and J. Maalampi.

\bibitem{VS} F. Vissani and A.Yu. Smirnov, hep-ph/9710565.

\bibitem{BP95}J.N. Bahcall and M.H. Pinsonneault, Rev. Mod. Phys.
{\bf 67}\,  (1995) 1.

\bibitem{KLP} P.I. Krastev, Q.Y. Liu and S.T. Petcov, Phys. Rev. {\bf
D 54} (1996) 7057.

\bibitem{BandG2} S. M. Bilenky, C. Giunti, Z. Phys. 
{\bf C 68} (1995) 495.


\bibitem{KrPe} P.I. Krastev and S.T. Petcov, private communications. 



\bibitem{BLS} K.S. Babu,  Q.Y. Liu, A.Yu. Smirnov, hep-ph/9707457,
updated November, 1997, to be published in  Phys. Rev. {\bf D}.



\bibitem{DN}
For recent updates, see e.g.,
Q.Y. Liu, M. Maris and S.T. Petcov, Phys. Rev. {\bf D 56},
 (1997) 5991; J.N. Bahcall, P.I. Krastev,
IASSNS-AST-97-31, Jun 1997, hep-ph/9706239; \\
 A.J. Baltz and J. Weneser, Phys. Rev. {\bf D 50} (1994) 5971.

\bibitem{MPet} M. Maris and S.T. Petcov, hep-ph/9803244.  


\bibitem{Qian} Y. -Z. Qian et al., {\sl Phys. Rev. Lett.},
{\bf 71} (1993) 1965; W.C. Haxton, K. Langanke, Y.Z. Qian,
P. Vogel, Phys. Rev. Lett. {\bf 78} (1997) 2694; Yong-Zhong Qian,
Nucl. Phys. {\bf A 621} (1997) 363c.   
\\
Y.Z. Qian, W.C. Haxton, K. Langanke, P. Vogel, Phys. Rev. {\bf C 55}
(1997) 1532. 


\bibitem{pelt} J. T. Peltoniemi, hep-ph/9511323. 


\bibitem{cris} {N. Hata et al.}, {\sl Phys. Rev. Lett.} {\bf  75} (1995)
3977.

\bibitem{sar} {P. J. Kernan and S. Sarkar}, {\sl Phys. Rev. }
{\bf D 54} (1996) 3681; {S. Sarkar}, {\sl Reports on Progress
in Physics} {\bf 59 } (1996) 1; Keith A. Olive, talk given at 5th
International Workshop on Topics in Astroparticle and Underground
Physics (TAUP 97), Gran Sasso, Italy, 7-11 Sep 1997; Keith A. Olive,
proceedings of 5th International Conference on Physics Beyond the 
Standard Model, Balholm, Norway, 29 Apr - 4 May 1997.  

\bibitem{ncris} { C. J. Copi, D. N. Schramm and M. S. Turner},
{\sl Phys. Rev. Lett.} {\bf 75} (1995) 3981;
{ K. A. Olive and G. Steigman,}  {\sl Phys. Lett.} {\bf B 354} (1995) 357.

\bibitem{Foot} R. Foot, R.R. Volkas, Phys. Rev. Lett. {\bf
75} (1995) 4350; Phys. Rev. {\bf D 55} (1997) 5147.  

\bibitem{Foot2} R. Foot,
M.J. Thomson, R.R. Volkas,
Phys. Rev. {\bf D 53} (1996) 5349. 

\bibitem{shi} X. Shi, Phys. Rev. {\bf D 54} (1996) 2753; X. Shi, 
D. N. Schramm,
B. D. Fields,
Phys. Rev. {\bf D 48} (1993) 2563.

 
\end{thebibliography}
\end{document}